# Quantum Information and Randomness

Johannes Kofler and Anton Zeilinger

Institute for Quantum Optics and Quantum Information (IQOQI), Austrian Academy of Sciences,
Boltzmanngasse 3, 1090 Vienna, Austria
Faculty of Physics, University of Vienna, Boltzmanngasse 5, 1090 Vienna, Austria

**Classical Physics and Determinism**

The desire to understand nature and the search for a causal explanation of all events have probably been among the driving forces in the evolution of humans since hundreds of thousands of years. Thus, it is no surprise that the philosophical debate on causality dates back to ancient times and many cultures, both east and west.

While scholars from different fields define causality often in different ways, modern natural science – starting in the 17th century with Galileo Galilei and Isaac Newton – established the clear framework of classical physics. There, the whole world follows not only causal but even *deterministic laws*. The significance of this worldview is best illustrated by Pierre-Simon Laplace's famous demon, a hypothetical entity which knows precisely the position and the velocity of any single particle in the universe as well as all the acting forces. By using Newton's laws it is then able to compute all the past and the future evolution of the world with absolute *certainty*. In 1814, Laplace wrote:[1]

> "We may regard the present state of the universe as the effect of its past and the cause of its future. An intellect which at a certain moment would know all forces that set nature in motion, and all positions of all items of which nature is composed, if this intellect were also vast enough to submit these data to analysis, it would embrace in a single formula the movements of the greatest bodies of the universe and those of the tiniest atom; for such an intellect nothing would be uncertain and the future just like the past would be present before its eyes."

By the beginning of the 20th century, the deterministic framework of classical physics, which had started from mechanics, also incorporated the theory of light, electricity and magnetism (electrodynamics), the theory of heat (thermodynamics and statistical physics), as well as the special and general theory of relativity about space and time that encompass also a modern post-Newtonian theory of gravitation.

But a small number of open problems at the end of the 19th century, such as the radiation from black bodies studied by Max Planck, turned out to be of vast importance. Slowly but steadily, experiment by experiment, the foundational structure of classical physics had come under attack. Albert Einstein had suggested a new heuristic aspect of the nature of light using particles in contradiction to waves. Niels Bohr had found phenomenological rules to describe the radiating electronic transitions in atoms which were in plain contradiction with classical electrodynamics. Louis de Broglie postulated a theory of matter waves, stating that not only light but also material particles like electrons have a dualistic nature and can behave as a wave or as a particle. By the middle of the 1930s the old classical world view had been given up, as Werner Heisenberg, Erwin Schrödinger, Max Born, Pascual Jordan, Wolfgang Pauli, Paul Dirac and John von Neumann had built a totally new theory – accurately describing all the new experimental findings – as one of the underlying foundational structures of physics: *quantum mechanics*. Today, quantum mechanics or, broader, quantum physics – though both terms are often used in a synonymous way – is a central pillar of many aspects of modern high tech industry including semiconductors, and therefore the inner workings of computers, and lasers, to just name two examples.

**Quantum Mechanics and Information**

Quantum mechanics in its Copenhagen interpretation was a paradigmatic shift in many ways. But most fundamentally, it *abandoned the assumption of an all-embracing objective reality*. In 1958, Heisenberg formulated this world view in the following way:[2]

> "[W]e are finally led to believe that the laws of nature that we formulate mathematically in quantum theory deal no longer with the particles themselves but with our knowledge of the elementary particles. [...] The conception of the objective reality of the particles has thus evaporated in a curious way, not into the fog of some new, obscure, or not yet understood reality concept, but into the transparent clarity of a mathematics that represents no longer the behavior of the elementary particles but rather our knowledge of this behavior."

Contrary to the prerequisite in classical physics, it is not true that objects posses definite properties at all times. In general, we can only make *probabilistic predictions*. In fact, the quantum mechanical *wave function* or *state* which is associated with a physical system is precisely a *catalog of information* about it and at the same time the complete (albeit probabilistic) description of all possible outcomes in future experiments.

An immediate consequence of this is *objective randomness*. From the viewpoint that we can in principle only make a probabilistic prediction of, say, the position of a particle in a future measurement, its precise position does not exist as an element of reality. The outcome in the measurement must be irreducibly random and non-deterministic. Causality is hence preserved only in a weak sense. There is of course still a dependence of results in a measurement on the previous preparation of a system. In other words, in quantum mechanics there is apparently still a lot of structure and order in the time evolution of systems, and not complete chaos and randomness. However, in general the evolution is such that the final individual measurement outcomes are not predetermined and that *only their probabilities are determined* by the theory. Causality in its deterministic reading – namely that the properties themselves are determined with certainty and that only one single outcome is possible at each measurement – is therefore given up.

Quantum randomness is connected with the *superposition principle*, i.e. the possibility that a system can be in a combination of a variety of different states. For example, the particle may be in a superposition of different places and therefore its position is not definite. In the famous double slit experiment, particles (e.g. photons or electrons) pass a wall with two slits, centered at $x_1$ and $x_2$ respectively. They do not have a classical trajectory though either one of the slits but they are in a superposition state of passing through the first slit centered at position $x_1$ *and* through the second slit centered at $x_2$, formally written as $(|x_1\rangle + |x_2\rangle)/\sqrt{2}$ (wave behavior). These possibilities later *interfere* on an observation screen, leading to an oscillating intensity pattern which, however, is built up from *individual scintillations* (particle behavior).

If *two or more systems* are in a superposition of different states, they are called *entangled*. For instance, one could prepare a pair of two particles, A and B, in a superposition of the state "particle A is at position $x_1$ and particle B is at position $x_3$" and the state "particle A is at position $x_2$ and particle B is at position $x_4$", formally written as $(|x_1\rangle_A |x_3\rangle_B + |x_2\rangle_A |x_4\rangle_B)/\sqrt{2}$. In such an entangled state, the composite system is completely specified in the sense that the *correlations* between the individuals are well defined. Whenever particle A is found at position $x_1$ (or $x_2$), particle B is certainly found at $x_3$ (or $x_4$ respectively). However, there is no information at all about whether particle A is at $x_1$ or $x_2$ and whether B is at $x_3$ or $x_4$. In 1935, in his article "The present situation in quantum mechanics" Schrödinger wrote:[3]

> "Total knowledge of a composite system does not necessarily include maximal knowledge of all its parts, not even when these are fully separated from each other and at the moment are not influencing each other at all."

**Local Realism and Bell's Theorem**

In a letter from 1926 to Born, Einstein expressed his deep conviction that the randomness of individual quantum mechanical events can always be reduced to an unknown cause with the words:[4]

> "I, at any rate, am convinced that He does not throw dice".

The answer attributed to Bohr was to suggest: "Einstein, don't tell God what to do." While Einstein's position can be seen as insisting that physics must answer ontological questions like "What is?", Bohr may be interpreted as limiting physics to answering epistemological questions like "What can be said?"

In 1935, Einstein, Boris Podolsky and Nathan Rosen (EPR) published the seminal article[5] "Can quantum-mechanical description of physical reality be considered complete?" In it, they used certain entangled states – now called "EPR states" – and exploited their "spooky" correlations to argue that quantum mechanics is an *incomplete* description of physical reality. As an analog of this position, one could say that also the probabilities in 19th century thermodynamics and statistical physics century simply arise just out of *ignorance* of the underlying complex microscopic situations, which are considered to be fully deterministic. Why should it not be possible that something very similar could be true for quantum mechanics?

In contrast to a closed world in which randomness is only subjective and due to ignorance of underlying causes, objective randomness manifests a genuinely open world. In a 1954, Pauli expressed his viewpoint in a letter to Born:[6]

> "Against all the retrogressive endeavors (Bohm, Schrödinger etc. and in some sense also Einstein) I am sure that the statistical character of the $\psi$-function and hence of nature's laws – on which you insisted from the very beginning against Schrödinger's resistance – will define the style of the laws at least for some centuries. It may be that later, e.g. in connection with the living processes, one will find something entirely new, but to dream of a way back, back to the classical style of Newton-Maxwell (and these are just dreams these gentlemen are giving themselves up to) seems to me hopeless, digressive, bad taste. And, we could add, it is not even a beautiful dream."

The search for an underlying theory was moved into the realm of experiments in the 1960s, when John Bell put forward his famous theorem. Bell made three assumptions about a possible physical worldview. 1. *Realism*: Objects posses their properties prior to and independent of observation. (Determinism is a specific form of realism, in which all properties have predefined values at all times.) This means there are "hidden variables" specifying the properties of systems. 2. *Locality*: This is the concept that if two systems no longer interact, no real change can take place in the second system in consequence of anything that may be done to the first system. 3. *Freedom of choice*: In Bell's words,[7] this hypothesis requires that the measurement settings chosen by the experimenters "can be considered as free or random" such that "they are not influenced by the hidden variables".

Bell mathematically proved that the combination of these three assumptions, denoted as *local realism*, is at variance with quantum mechanics. He derived an inequality, which is satisfied by all local realistic theories but which can be violated by entangled quantum states. Starting in the 1970s, a series of ground-breaking experiments disproved local realism and showed perfect agreement with the predictions from quantum theory. Until today, numerous experiments have been and are still performed, achieving better and better accuracy and closing more and more of the loopholes of earlier experiments. In view of the importance of this question, a loophole-free Bell test should be performed experimentally, even though already now it seems very unlikely that the concept of local realism – i.e. the concept of an objective external reality without any "spooky action at a distance" – can be maintained.

A decisive feature in experiments violating Bell's inequality is that entangled states of particle pairs are used. One particle of each pair is measured at one location by a party usually called Alice. The other particle is detected at some distant location by another party usually called Bob. Alice and Bob both independently and freely choose their measurement setting and note down the outcomes. Due to the entanglement of the initial state, the correlations between their outcomes are well defined. They are in a certain sense "stronger" than what is possible in any (classical) local realistic theory and thus violate Bell's inequality. But, according to the Copenhagen interpretation, Alice's and Bob's individual outcomes are objectively random and undefined in each run of the experiment until they manifest themselves in the measurement process. This is why there is no conflict with special relativity theory, although Alice and Bob might be far apart from each other and make their measurements at the same time. There is *no need to transmit physical information* from one side to the other. The resource for the quantum correlations is in the entangled state they share and not in any transmissions from one side to the other. It is the randomness (disliked by Einstein) which saves entanglement (disliked by Einstein) from violating special relativity (liked by Einstein).

It is worth noting that there have been suggestions where individual particles possess well defined properties (like position and momentum) at all times. De Broglie's pilot wave model from the 1920s implied a non-linear evolution in quantum mechanics. Based on this model, in the 1950s David Bohm developed his Bohmian interpretation of quantum mechanics, also often called the de Broglie-Bohm theory. In contrast to the Copenhagen interpretation, Bohmian mechanics saves the deterministic worldview by assuming that all particles have definite positions at all times (which are the hidden variables of the theory) and follow trajectories which are guided by a quantum potential given by the wave function. Randomness in this theory is indeed only *subjective*, stemming from ignorance of the hidden variables.

However, according to Bell's theorem, the price to pay is that this guiding potential *has to change non-locally* (instantaneously), if one has an entangled state between two distant parties and one party makes a measurement. The measurement on one particle changes the quantum potential in the whole space, in particular at the place of the second particle. This leads to a strong tension with the special theory of relativity. While the testable predictions of Bohmian mechanics are isomorphic to standard Copenhagen quantum mechanics, its underlying hidden variables have to be in principle unobservable. If one could observe them, one would be able to take advantage of that and signal faster than light, which – according to the special theory of relativity – leads to physical temporal paradoxes.

**Quantum Information and Technology**

In the last decades, it has turned out that foundational aspects of quantum physics are not only of philosophical interest but have opened up striking new possibilities of a new quantum information technology. Features like superposition and entanglement can be exploited to solve certain tasks which cannot be solved – or at least cannot be solved that efficiently – by purely classical machines. Among the most famous developments in this new field of quantum information, i.e. the union of quantum theory with information theory, are *quantum computation* and *quantum cryptography*.

In quantum computation, one exploits the entanglement of quantum systems as well as specific operations like transformations or measurements to find the result of certain problems in fewer steps and thus faster than any classical computer can possibly do. In one particular realization of quantum cryptography, one sends pairs of entangled particles to two parties, Alice and Bob. Due to the quantum correlations, Alice and Bob can establish a secret key. Any eavesdropping can be revealed by checking a Bell's inequality. As long as it is violated, no eavesdropper could have possibly intercepted the particles and hence know the key. While quantum computation is still at the stage of fundamental research and limited to only a few quantum bits ("qubits") and simple calculations, quantum cryptography has already made the step to an industrial level.

Another task which relies on the laws of quantum theory is the *teleportation of an unknown quantum state*. There, Alice and Bob share an entangled particle pair: particles 2 and 3 in Figure 1. Alice is provided with another particle, particle 1 in the Figure, which is in an arbitrary input state. Alice then performs a joint measurement on particles 1 and 2 in the so-called Bell-basis, i.e. she projects her two particles on a maximally entangled state. She then sends her result to Bob, e.g. via telephone or internet. Depending on this information, Bob performs one out of four simple transformations on his particle 3. Finally, due to the initial entanglement, the state of Bob's particle 3 is identical to the initial input state of Alice's particle 1, which has lost its private properties.

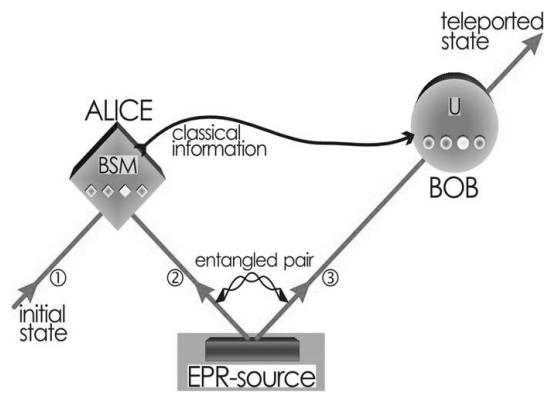

Figure 1: The scheme of quantum teleportation. Alice and Bob share an entangled state which is produced by an EPR source. Alice performs a joint Bell state measurement (BSM) on her particle and on an input particle. She transmits her result to Bob who, depending on Alice's message, performs an operation on his particle. The final state of Bob's particle is then identical to Alice's input state.[8]

Which of the four possible outcomes Alice gets, is completely random. She has to transmit this information to Bob via a classical communication channel. Bob has to wait for this information to perform the correct operation to his particle. Therefore, due to the randomness of the Bell state measurement, there is no conflict with special relativity. After all, no information is transmitted faster than light. If Alice could somehow break the complete randomness such that she would be able to change the statistics at Bob's location, then she could communicate faster than the speed of light.

**A Foundational Principle for Quantum Mechanics**

It is worth asking, whether there is a deeper message conveyed by the Copenhagen interpretation of quantum mechanics. Can one come up with a *foundational principle* for quantum mechanics which can explain some of its key features like randomness and entanglement?

In this regard, in 1999, one of us (A.Z.) has put forward an idea which connects the concept of information with the notion of elementary systems. For the subsequent line of thought, we first have to make ourselves aware of the fact that our description of the physical world is represented by *propositions*, i.e. by logical statements about it. These propositions concern classical measurement results. Therefore, the measurement results must be irreducible primitives of any interpretation. And second, that we have knowledge or information about an object only through observations, i.e. by interrogating nature through yes-no questions. It does not make any sense to talk about reality without the information about it.

Any complex object which is represented by numerous propositions can be decomposed into constituent abstract systems that each need fewer propositions to be specified. The position of a particle, for instance, can be decomposed into a sequence of "left or right" questions, subdividing space into smaller and smaller parts. The process of subdividing reaches its limit when the individual subsystems only represents a single proposition, and such a system is denoted as an *elementary system*. This notion is very closely related to Carl Friedrich von Weizsäcker's binary alternative, called the "Ur",[9] and to the qubit of modern quantum information theory. The truth value of a single proposition about an elementary system can be represented by one bit of information with "true" being identified with the bit value "1" and "false" with "0".

In "A foundational principle of quantum mechanics"[10] and a subsequent work[11] it is then suggested to assume a *principle of quantization of information*:

*An elementary system is the manifestation of one bit of information.*

Disregarding the mass, charge, position and momentum of an electron, its *spin* is such an elementary system. Now consider that we prepare many electrons, each in such a spin state that it is certainly deflected into the positive vertical, i.e. the $+z$, direction if it passes a Stern-Gerlach magnet which is also oriented along $+z$ (see Figure 2).

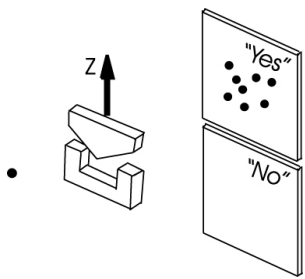

Figure 2: Each individual electron can be prepared in such a quantum spin state that it is deflected in the $+z$ direction when it passes the inhomogeneous magnetic field of a Stern-Gerlach magnet oriented also along $+z$. The question "spin up along $z$?" is hence answered with "yes" by all the electrons.[11]

The corresponding quantum state of each of these electrons is denoted by $|+z\rangle$, and it specifies the truth value "true" or "1" of the 1-bit proposition

*"The spin is 'up' along the z direction."*

Now we have used up our single bit per electron spin. Therefore, a measurement of an electron along any other direction, say along the direction $\theta$ must necessarily contain an element of randomness (see Figure 3). This randomness must be objective and irreducible. It cannot be reduced to unknown hidden properties as then the system would represent more than a single bit of information. Since there are more possible experimental questions than the system can answer definitely, it has to "guess". Objective randomness is a *consequence* of the principle lack of information.

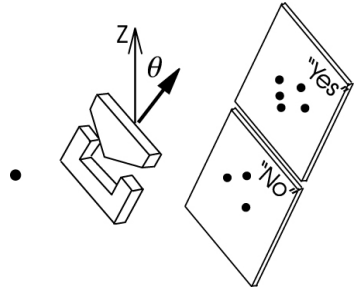

Figure 3: If electrons are again prepared with spin up along *z*, but are measured along a tilted direction *θ*, some of them are deflected in the *θ*-direction and some in the opposite direction. For each individual event, only the probability is determined by quantum theory. The question "spin up along *θ*?" is answered with "yes" by some of the electrons and with "no" by the others.[11]

In the extreme case when the measurement direction is orthogonal to the preparation, say along *y*, the system does not carry any information whatsoever about the measurement result. Due to this full *complementarity* the outcomes are completely random with probability ½ each. After the measurement, however, the system is found to be in one of the states "up along *y*" or "down along *y*". This new one-bit information was spontaneously created in the measurement, while the old information is irrecoverably erased due to the limitation of one bit "storage capacity" in the electron spin.

Let us now proceed to multi-particle systems. First, the above information principle is generalized to

*N elementary systems are the manifestation of N bits of information.*

Let us start with the simple example of a product state of two spin particles, labeled 1 and 2: $|+z\rangle_1|-z\rangle_2$. The two bits of these two elementary systems represent the truth values of the two following propositions:

"The first spin is up along *z*." &
"The second spin is down along *z*."

The two bits have thus been used up to define *individual properties* of the two electrons. But we can also choose to use the two bits for defining *joint properties* of the two spin. For instance we may specify the truth of the following two propositions:

"The two spins are different along z". &
"The two spins are the same along y".

The corresponding quantum state is uniquely defined. It is the entangled (EPR-like) state

$$|\text{EPR}\rangle_{12} = \tfrac{1}{\sqrt{2}}\left(|+z\rangle_1|-z\rangle_2 + |-z\rangle_1|+z\rangle_2\right)$$
$$= \tfrac{1}{\sqrt{2}}\left(|+y\rangle_1|+y\rangle_2 - |-y\rangle_1|-y\rangle_2\right)$$

Given the principle of quantization of information, entanglement is therefore a *consequence* of the fact that the total information is used to define joint and not individual properties of a composite system. While, due it its quantum correlations, this entangled state is capable of violating local realism in a Bell test, it does neither specify whether the individual spins are up or down along *z* nor whether they are up or down along *y*. The individual properties remain completely undefined.

In realistic (deterministic) non-local theories like Bohmian mechanics, every particle carries a huge amount of information, specifying the results of all possible measurement outcomes. However, being in accordance with the Copenhagen interpretation, the foundational principle of finiteness of information represented by elementary systems provides a new way of *understanding* objective randomness, complementarity and entanglement. Recently, the approach discussed here has also been connected with the notion of logical independence in pure mathematics.[12]

**Conclusion**

In retrospect, regarding the notion of causality, the framework of physics has undergone a paradigmatic change with the advent of quantum mechanics about 80 years ago. The deterministic character of physics has been abandoned and knowledge and information have become central concepts. The foundations of quantum theory have not only shone new light on one of the deepest philosophical questions, namely the nature of reality, but have in the past decades also led to the possibility of new technologies.

**Acknowledgements**

We acknowledge financial support by the Austrian Science Fund, the European Research Council, and the Templeton Foundation.